\documentclass[twocolumn, superscriptaddress,secnumarabic,amssymb, nobibnotes, prl,aps,showpacs,10pt]{revtex4-1}
\usepackage{graphicx}
\usepackage[integrals]{wasysym} 
\usepackage{nicefrac,amssymb,amsmath} 

\newcommand{\nrho}{n}
\def\vec#1{{\mathbf{ #1}}}
\begin{document}
\begin{abstract}

The charge-density response of extended materials is usually dominated by the collective oscillation of electrons, the plasmons. Beyond this feature, however, intriguing many-body effects are observed. They cannot be described by
one of the most widely used approaches for the calculation of dielectric functions, which is time-dependent density functional theory (TDDFT) in the adiabatic local density approximation (ALDA). 
Here we propose an approximation to the TDDFT exchange-correlation kernel which is non-adiabatic and non-local. It is derived in the homogeneous electron gas and implemented in the real system in a simple mean density approximation. This kernel contains effects that are completely absent in the ALDA; in particular, it correctly describes the double plasmon in the dynamic structure factor of sodium, and it shows the characteristic low-energy peak that appears in systems with low electronic density. It also leads to an overall quantitative improvement of spectra. 

\end{abstract}

\title{Non-local and non-adiabatic effects in the charge-density response of solids: a time-dependent density functional approach}

\newcommand{\lsi}{Laboratoire des Solides Irradi\'es, \'Ecole Polytechnique, CNRS, CEA,  Universit\'e Paris-Saclay, F-91128 Palaiseau, France}
\newcommand{\etsf}{European Theoretical Spectroscopy Facility (ETSF)}
\newcommand{\soleil}{Synchrotron SOLEIL, L'Orme des Merisiers, Saint-Aubin, BP 48, F-91192 Gif-sur-Yvette, France}

\author{Martin Panholzer}%
\affiliation{\lsi}
\affiliation{\etsf}

\author{Matteo Gatti}
\affiliation{\lsi}
\affiliation{\etsf}
\affiliation{\soleil}

\author{Lucia Reining}
\affiliation{\lsi}
\affiliation{\etsf}

\pacs{71.10.-w, 71.15.-m, 71.45.Gm, 78.20.Bh}
\maketitle


The response to an external perturbation is an important tool to probe materials, and spectroscopic experiments play a crucial role for their understanding. Response properties are often of interest for applications: examples are the linear response to photons, which governs optical properties and hence the color of materials and their capability to absorb the sunlight, or the response to a beam of fast charges, which determines the stopping power. 

A first attempt to interpret experimental findings or to predict response properties is based on the band structure, in a simple independent-particle picture. However, collective effects and signatures of strong correlation influence and sometimes even dominate electronic spectra, making their calculation a formidable intellectual challenge and a crucial tool for technological applications. 
This is immediately apparent in the case of optical absorption spectra, where the Coulomb interaction can lead to bound electron-hole pairs that create sharp excitonic peaks in the fundamental gap. Other important spectroscopic quantities are the loss function and the dynamic structure factor as measured in electron energy loss spectroscopy or inelastic x-ray scattering \cite{Schulke2007}; for example, the loss function  exhibits the plasmon excitations and is therefore a key ingredient for plasmonics. It is also crucial for theory, since it yields the screened Coulomb interaction, which enters the calculation of the correlation energy in the adiabatic connection formula \cite{Gunnarsson1976,Langreth1977} and which is one of the main building blocks of many-body perturbation theory \cite{Martin2016}. Its calculation is at first sight straightforward, since the spectra are often dominated by classical electrostatic (Hartree) effects for which the Random Phase Approximation (RPA) is sufficient to capture the essential trends \cite{Schulke2007}. Beyond the RPA, the adiabatic local density approximation (ALDA) to time-dependent density functional theory (TDDFT) yields in general a small quantitative improvement \cite{Botti2007}. However, for today's needs
often the resulting rough overall agreement is not sufficient.
Details of the loss function  are responsible for the shape of the satellite spectra in photoemission, and transmit therefore precious information, for example about doping \cite{Verdi2017}; this holds even in simple semiconductors. Especially in correlated materials, low energy structures can dramatically influence materials properties, even when they are very weak \cite{Rodl2017}.
Loss spectra can exhibit many-body effects such as lifetime broadening \cite{Rahman1984,Weissker} or double-plasmon excitations \cite{Sternemann2005,Huotari2008}, and in the low-density regime the spectral shape can be very different from the naively expected single plasmon peak, even in the homogeneous electron gas (HEG) \cite{Takada2016}. To capture those intriguing effects, one has to go beyond the simple approximations. 
Indeed, many advanced explicit or implicit density functionals have been developed that can be directly applied to real materials, some of them with the desired non-local behaviour \cite{Petersilka1996,Corradini1998,Nazarov2010,Trevisanutto2013}, others which are non-adiabatic \cite{Gross1985,Conti1997,Qian2005,Constantin2007,Nazarov2009}. However, most of them are meant to improve just one of the shortcomings of the ALDA; in particular, long-range corrected kernels are often adiabatic \cite{Reining2002,Botti2004,Sharma2011,Rigamonti2015}. Interestingly, even outside density functionals dynamical effects are difficult to capture; in particular the widely used Bethe-Salpeter equation within many-body perturbation Green's function theory is usually applied in an adiabatic approximation \cite{Onida2002}. On the other side, advanced calculations in the TDDFT framework exist in the HEG 
\cite{Takada2016}, but without an indication of how to use the expressions in a real material. 


The aim of the present work is to close this gap. In the spirit of ground-state density functionals such as the LDA \cite{Kohn1965}, our procedure is divided into two parts: first, an advanced calculation of the exchange-correlation kernel in the HEG, 
and second, a ''\textit{connector}'', namely a prescription of how to use the result in real materials. The enormous advantage of such a procedure is that the advanced calculation is done only once; indeed our HEG results are freely available\cite{2h2ptable}. By using the here proposed very simple connector, results for real materials are then obtained with an additional computational effort that is negligible with respect to the RPA. They show features that are extremely difficult to obtain otherwise, in particular double plasmon excitations and the double-peak structure which characterizes strong correlation in the low-density regime.

\begin{figure}[tb]
 \includegraphics[width=8.6cm,keepaspectratio=true]{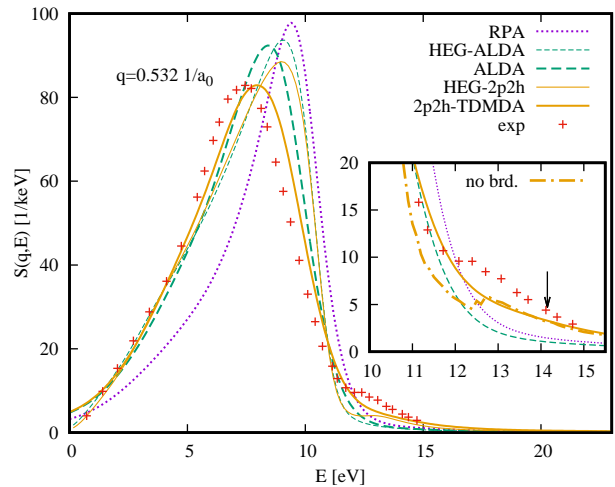}
 \caption{The dynamic structure factor of Sodium at $q=0.53\; 1/a_0$: experimental IXS results of ref. \cite{Cazzaniga2011} (red crosses) and different levels of theory for the real material (thick) and for the HEG at $r_s = 3.9x$ (thin), namely RPA (purple, short dashed), ALDA (green, dashed) and 2p2h (magenta, full). Inset: zoom on 
the double plasmon, in particular 2p2h (yellow) with (full) and without (dot-dashed) Gaussian broadening. Arrow: energy of the double plasmon in ref. \cite{Sternemann2005}.}
 \label{fig:Na_inhom}
\end{figure}

For an illustration we concentrate on a prototype material which is close to the HEG 
and which shows several interesting features, namely sodium. Fig. \ref{fig:Na_inhom} shows experimental \cite{Cazzaniga2011,Huotari2011} and various theoretical results for a moderate momentum transfer. In agreement with existing literature \cite{Cazzaniga2011,Huotari2011}, the dynamic structure factor obtained in the RPA exhibits one plasmon peak that is blue shifted with respect to experiment. The ALDA moves spectral weight to lower energies. However, the plasmon energy is still too large compared to experiment, and the peak is too asymmetric. Moreover, the double plasmon, which is clearly visible in experiment around 12 eV, is completely absent in the ALDA. In order to improve the spectral shape, one can add quasiparticle lifetime corrections \cite{Cazzaniga2011,Huotari2011}, but this is a quite empirical combination of density- and Green's functions-functionals, and it can moreover not access the double plasmon. The latter has been calculated diagramatically  in the HEG by Sturm and Gusarov (SG) \cite{Sturm2000}, and qualitative agreement with experiment was found for this feature \cite{Sternemann2005}. A finer comparison is hindered by the fact that even in sodium the effects of the crystal are not negligible: the ALDA result for the HEG given in Fig. \ref{fig:Na_inhom} shows a significant difference from the ALDA in real sodium. In order to distinguish effects of the crystal from the description of many-body effects,  let us first look at the HEG. 

The first task is to calculate the density-density response function $\chi(q,\omega,\nrho )$, which is a function of wavevector $q$, frequency $\omega$ and the homogeneous density $\nrho$. For our purpose this must be done on a level of theory that includes correlations and the explicit coupling of excitations sufficiently well. A suitable starting point is the correlated equations of motion approach of B\"ohm {\it et al.} \cite{Bohm2010}. It relies on a Jastrow correlated ground state $\bigl| \psi_0 \bigr\rangle=F \bigl| \phi_0 \bigr\rangle / \cal{N} $, where $\phi_0$ is a Hartree-Fock ground state and $\cal{N}$ is the normalization. $F$ is the correlation operator. When restricting it to two particle correlations 
it reads $F= e^{\frac{1}{2}\sum_{i<j} u(\vec{r}_i-\vec{r}_j)} $. The correlation functions $u$ are found by minimization of the energy.  
Excited states are described by neutral excitations of $\bigl| \phi_0 \bigr\rangle$, while $F$ is kept constant. 
In this work single particle hole and two-particle two-hole excitations are included. These excitations are optimized by employing the least-action principle. Compared to the original work \cite{Bohm2010} we use here a more complete description of the pair propagator. This is important to obtain the correct $q\rightarrow 0$ behavior of the plasmon dispersion in the pair propagator. Details of the derivation are in the supplementary material\cite{suppmat}.


For the calculation of $\chi$ we need matrix elements involving the correlated excited states. This is done 
within the correlated basis functions formalism \cite{CBF_1,CBF_2}. It turns out that the static structure function $S(q)$ appears as a fundamental ingredient in the result, while the explicit knowledge of $u$ is not needed. We can therefore profit from Monte Carlo (MC) calculations, which yield $S(q)$ with high precision \cite{Gori-Giorgi2000}, whereas they do in general not access spectra. 

\begin{figure}[tb]
 \includegraphics[width=8.6cm,keepaspectratio=true]{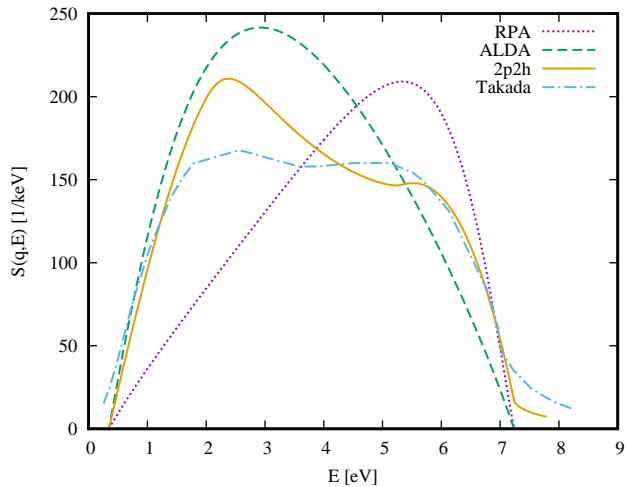}
 \caption{Dynamic structure factor of the HEG for $r_s=8$ at $q=2.2\; k_F$. 2p2h results (yellow continuous) are compared with RPA (purple short-dashed), TDLDA (green long-dashed), and with the results of Takada \cite{Takada2016} (blue dot-dashed).}
 \label{fig:rs8}
\end{figure}

A typical $\chi(q,\omega,n)$ calculated in this way for q=2.2 $k_F$ and a density corresponding to $r_s=8$ is shown in Fig. \ref{fig:rs8}. The RPA shows a broad and featureless plasmon excitation around 5.5 eV. This can be compared to the more realistic result of Takada \cite{Takada2016}, which has been obtained with an improved version of the Richardson Ashcroft exchange-correlation kernel \cite{Richardson1994}. This kernel, which satisfies many exact conditions, is wavevector- and frequency dependent. However, it has been derived for imaginary frequencies, and while it has been shown to yield significant improvement in correcting RPA correlation energies \cite{Lein2000}, its performance for spectroscopy is less obvious. Indeed, \textit{improvement by hand by Takada, Vignale?}. We thus expect that the corrections with respect to the RPA are significant, while there may still be deviations from the (unknown) exact spectrum, in particular no double plasmon.
In any case, Takada's result represents probably today the best available benchmark. It is remarkably different from the RPA: there is a strong shift of spectral weight to lower energies, and a double peak structure appears.  
The shift of spectral weight is at the origin of the so-called ``ghost plasmon'' 
(called ``ghost exciton'' in Ref. \cite{Takada2016}), which has been introduced by Takayanagi and Lipparini \cite{Takayanagi1997} as a pole of the \textit{irreducible} response function on the imaginary axis. It manifests itself as a negative static dielectric function, hence an attractive screened Coulomb interaction. Such as feature is always absent in the RPA. However, interestingly it is present in the ALDA, which indeed leads to a strong redshift of spectral weight, as can be seen in Fig. \ref{fig:rs8}. 
This can be understood, as in the $q\rightarrow 0$ limit ALDA approaches the MC result. Consequently the onset density for the appearance of this ghost plasmon is the same as in the MC prediction.
The link between the spectral weight shift and the sign of the screened Coulomb interaction $W$ can be understood from the spectral representation of the latter, which reads
$ 
W = v_c + \sum_s \frac{\omega_s W^s}{\omega^2 - \omega_s^2},
$ 
Where $v_c$ is the bare interaction, $\omega_s$ are discrete or continuous excitation energies, and $W^s$ are their squared amplitudes. At $\omega=0$ this expression becomes negative when low energy modes are sufficiently enhanced. However, as compared to ref. \cite{Takada2016} the ALDA does not create this effect in the correct way, since there is an overall shift rather than the appearance of a low energy mode in addition the original plasmon, which should lose intensity but change its position only very weakly. 
Our 2p2h results, instead, yield results that are close to Takada's ones in both respects: the shift of oscillator strength, and a (in our result slightly more intense) distinct low energy mode, which is to a large extent responsible for the shift of oscillator strength and therefore, for the appearance of the ghost plasmon.

The low energy mode, which can be clearly seen around  $q=2k_F$ for relatively  high $r_s$ values, has been introduced by Takada as an  ``excitonic collective mode''. From an extrapolation of his results up to $r_s=22$, he concludes that it
falls always inside the single-pair excitation region. Our approach leads to a slightly different conclusion. 
Indeed, it has been shown theoretically, with the same approach as used here, and confirmed experimentally 
that this mode can exist below the particle--hole continuum for 2D $^3$He by Godfrin et al. \cite{nature}. 
The strength of the mode indicates the proximity of the system to a phase transition to the solid (i.e. the Wigner solid for electrons) and this mode becomes the soft mode of this phase transition\cite{Nozières2004}. 
The driving force for these quantum phase transitions in is the same in both systems, i.e. the increasing ratio of the interaction over typical excitation energies (which is increased by compression in He and by expansion for electrons) and therefore correlations. Even away from the phase transition, however, as discussed above low energy modes play an important role in the spectroscopy of correlated systems, and the fact that our approach is able to reproduce this phenomenon is an important check. 

Another feature of interest is the double plasmon, which can by definition only be obtained with a frequency-dependent kernel. 
The result of our calculations, performed at the density of sodium ($r_s=4$) and at a wavevector of $q=0.53 1/a_0$, is shown in Fig. \ref{fig:Na_inhom}. With respect to the RPA, there is a visible enhancement of oscillator strength in the region of interest with a structure around 12.8 eV, which can be attributed to the double plasmon. This is close to the experimental results of  Huotari {\it et al.} \cite{Huotari2008} at 12.8 eV, or more recent results \cite{Cazzaniga2011,Huotari2011} which suggest 12.5 eV (see inset in Fig. \ref{fig:Na_inhom}). To say more about the significance of this result, in view of the non negligible differences between sodium and the HEG, we have to move closer to the real system, for which none of the above mentioned approaches has yet been applied. 

Our strategy takes advantage of the Dyson-like linear response equation for $\chi$, which for the HEG can be written as
\begin{equation}
  f^{hom}_{\rm xc}(q,\omega,\rho)=\frac{1}{\chi_0(q,\omega,\rho)}-\frac{1}{\chi(q,\omega,\rho)}-v_c(q) \;.
  \label{eq:fxc-hom}
\end{equation} 
 Here $v_c$ is the Coulomb interaction. The Lindhard function $\chi_0$ contains the information about the non-interacting system, whereas the exchange-correlation kernel describes interaction effects. Of course, it also depends on the system, which in the case of the HEG manifests itself by the fact that $f^{hom}_{\rm xc}$ is a function of the density. Still, focusing on  $f^{hom}_{\rm xc}$ instead of the full $\chi$ allows us to separate materials- and interaction-effects to a significant extent.  
Starting from our advanced results for $\chi$, we therefore calculate $f^{hom}_{\rm xc}$ according to Eq. \ref{eq:fxc-hom}. 



Only the $q\rightarrow 0$ limit is delicate because of lack of precision in the fitting procedure of the MC data. The easiest solution is to correct this limit using the static $f^{Corradini}_{\rm xc}(q)$ of 
Corradini {\it et al.} \cite{Corradini1998}, which should be close to exact in the static limit \cite{suppmat}. 




The such created table of $f_{\rm xc}^{hom}$ \cite{2h2ptable} has now to be used in the real systems: we have to devise the \textit{connector}. In the case of ground state calculations the canonical starting point is the LDA, based on the nearsightedness principle \cite{Kohn1996}. 
However, the baseline in the construction of $v_{\rm xc}({\bf r})$ is mostly to take some suitable density average around the local point ${\bf r}$. In the present case of $f_{\rm xc}({\bf r},{\bf r}';\omega)$ at least two regions (around ${\bf r}$ and around ${\bf r}'$), if not a whole area indicated by these two points, should give important contributions. If the pertinent area is larger than the scale of the density variation in the system, the canonical approximation should be to take the mean density $\overline \nrho$ of the system, rather than a somehow defined local density. Since a wavevector $q\approx 0.5/a_0$ as used in Fig. \ref{fig:Na_inhom} would roughly correspond to a distance of $2\pi/q = 4\pi a_0$, which is already large compared to the interatomic distance of $7 a_0$ in Na, we propose to adopt the \textit{mean density approximation} (MDA) as simplest connector to import $f_{\rm xc}$ from the HEG into extended systems \footnote{Of course, in the case of finite systems this idea would have to be adapted}. In practice, this means that in the real system one has to solve the Dyson equation for $\chi$ using the inhomogeneous $\chi_0$ and $f^{hom}_{\rm xc}$ calculated at $\overline \nrho$ of the real system. This is the time-dependent MDA (TDMDA). 

\begin{figure}[tb]
 \includegraphics[width=8.6cm,keepaspectratio=true]{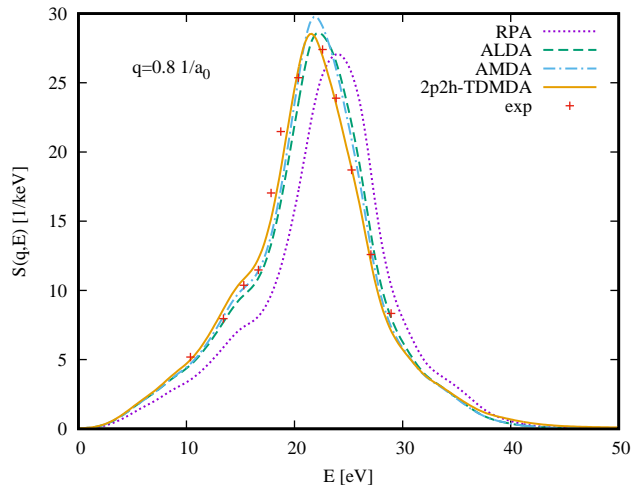}
 \caption{The dynamic structure factor of Silicon at $q=0.8\; 1/a_0$ compared with IXS results of Weissker {\it et al.} \cite{Weissker}. Lines have the same meaning as in Fig. \ref{fig:Na_6}, but additionally the TDMDA result is shown (blue).}
 \label{fig:Si_08}
\end{figure}

In order to get a feeling whether the choice between some local and a mean density is very delicate, we have calculated spectra for several systems using either the ALDA or the TDMDA. Here,  for comparison the TDMDA imports from the HEG the same static and local $\delta({\bf r}-{\bf r}')f^{hom}_{\rm xc}(q\to 0;\omega =0)$ as does the ALDA, but at the mean, instead of the local, density. For sodium (not shown) the spectra are on top of each other, which is not surprising. More interesting is silicon: it turns out that ALDA and TDMDA are still very close, as can be seen in Fig. \ref{fig:Si_08}. If there is any difference, the tendency is rather in favor of the TDMDA, which is confirmed by results in the supplemental material. This confirms that the TDMDA is a at least reasonable approximation, and we will adopt it in the following.




\begin{figure}[tb]
 \includegraphics[width=8.6cm]{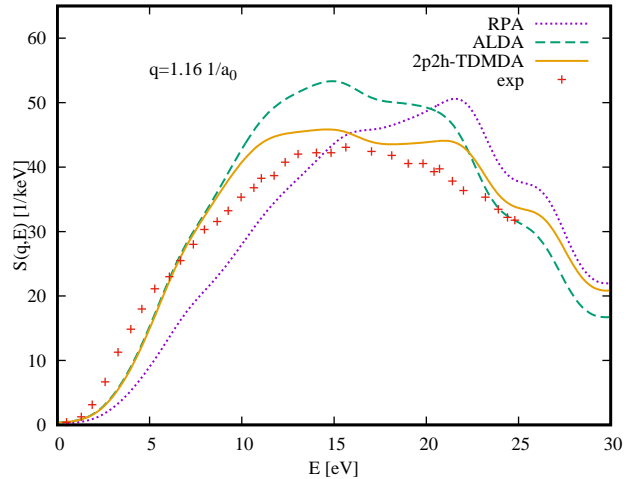}
 \caption{Similar to Fig. \ref{fig:Na_inhom}, at higher momentum transfer $q=1.16\; 1/a_0$. The homogeneous results are omitted here.}
 \label{fig:Na_6}
\end{figure}

With the TDMDA, any whatsoever sophisticated interpolated or tabulated HEG kernel is easy to import and to use in real systems. This allows us to calculate again sodium, now using our $2p2h$ kernel. For the ground state calculation leading to $\chi_0$, we adopt the LDA. The mean density is that of the valence electrons, which are well separated from the core electrons. The result for $q=0.532 a_0^{-1}$ is given in fig. \ref{fig:Na_inhom}. The difference with respect to the $2p2h$ homogeneous result is mainly a significant redshift of the main plasmon. Of all the approximations shown, the $2p2h$-TDMDA result has the best  agreement with experiment. Since the double plasmon is only accessible by a frequency-dependent kernel, it merits particular attention; therefore  the inset in fig. \ref{fig:Na_inhom} shows a zoom. 
The double plasmon contribution is clearly visible in the experiment and the theory, especially when we remove the gaussian broadening. (This is necessary, since plasmon peak position in the 2p2h result is still a bit higher in energy and the broadening buries the double plasmon partly.)
The theoretical position is close to the homogeneous result and to experiment, and improves over the homogeneous result of Sturm and Gusarov by more than 1 eV.  

In order to probe also the impact of the wavevector-dependence of the $2p2h$-TDMDA kernel, fig. \ref{fig:Na_6} shows results for a larger wavevector, $q=1.16 1/a_0$. The $2p2h$-TDMDA result is again better than the ALDA; \emph{ in this now larger energy range the absence of lifetime damping in the ALDA becomes visible, which is cured by the self-energy diagrams in the $2p2h$ approach. I(Martin) don't understand this} Similar improvement is obtained for silicon, as shown in the supplemental material. 


In conclusion, we have shown that the nonlocal and dynamical   2p2h kernel can be  accurately built in the HEG and successfully used for the calculation of dynamic structure factors in real solids. 
The new kernel improves over existing approximations for the HEG, producing at the same time double-plasmon and ``ghost-exciton'' features 
that, so far, have been separately obtained through distinct approximations only \cite{Takada2016,Sturm2000}. 
Moreover, the 2p2h kernel can be imported in real materials through a simple connector that is based on the mean electronic density. This severe approximation precludes its use in the present form in finite systems, and it cannot describe band gaps or bound excitons in insulators \cite{Reining2002,Botti2007}. However, it significantly improves the dynamic structure factor of simple metals and semiconductors, including double-plasmon resonances which are completely missed by standard TDDFT approximations. This implies that it could in principle also be used to improve many-body perturbation theory based on the screened Coulomb interaction, or total energy calculations using the adiabatic connection \cite{Gunnarsson1976,Langreth1977}. 
Finally, we advocate that the very good quality of these results illustrates a general strategy that is very promising, as it allows one to separate dynamical correlation effects, which  can be calculated and tabulated once for all in the HEG \cite{2h2ptable}, from electronic structure features that are material specific but in principle easier to deal with.

\bibliographystyle{apsrev4-1} 
\bibliography{lib}
\end{document}